
\input phyzzx
\hoffset-.3cm
\pubnum{IC/93/144}
\date{June, 1993}
\titlepage
\title{
Thermodynamics  and Form Factors of Supersymmetric Integrable Field
Theories}
\author{Changrim Ahn}
\address{  International Centre for Theoretical Physics\break
                  Strada Costiera 11\break
                  34014 Trieste, ITALY}
\abstract{
\noindent We study on-shell and off-shell properties of
the supersymmetric sinh-Gordon and perturbed SUSY Yang-Lee
models using the thermodynamic Bethe
ansatz and form factors.
Identifying the supersymmetric models with the Eight Vertex
Free Fermion Model,
we derive inversion relation for inhomogeneous transfer
matrix and TBA equations and get correct UV results.
We obtain two-point form factors of the trace of energy-momentum
tensor using the Watson equations and their SUSY transformations.
As an application, we compute the UV central charge
using these form factors and spectral representation of the $C$-theorem.
}
\def\P{\Phi}
\def\D{\Delta}
\def\G{\Gamma}
\def\T{\Theta}

\def\a{\alpha}
\def\b{\beta}
\def\g{\gamma}
\def\d{\delta}
\def\ep{\epsilon}
\def\s{\sigma}
\def\p{\phi}
\def\t{\theta}
\def\l{\lambda}
\def\L{\Lambda}

\def\m{\mu}

\def\r{\rho}
\def\half{{1\over{2}}}
\def\pd{\partial}
\def\sl#1{\not{\hbox{\kern-2pt ${#1}$}}}
\def\wh#1{\widehat{#1}}

\def\ol#1{{\overline#1}}
\def\cd{\cdot}


\def\ket#1{{ \big\vert{#1}\bigr\rangle }}
\def\bra#1{{ \bigl\langle {#1}\big\vert }}

\REF\ZamZamol{A. B. Zamolodchikov and Al. B. Zamolodchikov, Annals. Phys.
{\bf 120} (1979)  253}
\REF\BPZ{A. A. Belavin, A. M. Polyakov, and A. B. Zamolodchikov,
Nucl. Phys. {\bf B241} (1984) 333}
\REF\Zamoli{Al. B. Zamolodchikov, Nucl. Phys. {\bf B342} (1990) 695}
\REF\Zamiii{A. B. Zamolodchikov, Int. Journ. of Mod. Phys. {\bf A4} (1989)
4235}
\REF\KarWei{M. Karowski and P.H. Weisz, Nucl. Phys. {\bf B139} (1978) 455;
M. Karowski, Phys. Rep. {\bf 49} (1979) 229}
\REF\Smibook{F. A. Smirnov, Form Factors in Completely Integrable Models of
Quantum Field Theory (World Scientific, Singapore, 1992)
and references therein.}
\REF\YurZam{V.P. Yurov and Al.B. Zamolodchikov, Int. J. Mod. Phys. {\bf A4}
(1989) 3419}
\REF\Zamolii{Al. B. Zamolodchikov, Nucl. Phys. {\bf B348} (1991) 619}
\REF\FMS{A. Fring, G. Mussardo and P. Simonetti,
Nucl. Phys. {\bf B393} (1993) 413;
A. Koubek and G. Mussardo, SISSA preprint, SISSA preprint ISAS-93-42}
\REF\CarMus{J.L. Cardy and G. Mussardo, UCSB and SISSA preprint
UCSBTH-93-12, ISAS-93-75}
\REF\BabBer{O. Babelon and D. Bernard, Phys. Lett. {\bf B288} (1992) 113}
\REF\DesDeV{C. Destri and H.J. De Vega, Nucl. Phys. {\bf B338} (1991) 251}
\REF\FLV{D.Z. Freedman, J. Latorre and X. Vilasis, Mod. Phys. Lett.
{\bf A6} (1991)  531}
\REF\Baxbook{R. J. Baxter, Exactly solved models in statistical mechanics
(Academic Press, New York, 1982)}
\REF\ABF{G. E. Andrew, R. J. Baxter, and P. J. Forrester, J. Stat. Phys.
{\bf 35} (1984) 193}
\REF\LeC{A. LeClair, Phys. Lett. {\bf 230B} (1989) 103; D. Bernard and
A. LeClair, Nucl. Phys. {\bf B340} (1990) 721}
\REF\ResSmi{F. A. Smirnov, Int. J. Mod. Phys. {\bf A4} (1989) 4213;
N. Yu Reshetikhin and F. Smirnov, Comm. Math. Phys. {\bf 131}
(1990) 157}
\REF\BerLeCii{D. Bernard and A. LeClair, Comm. Math. Phys. {\bf 142} (1991)
99}
\REF\ABL{C. Ahn, D. Bernard and A. LeClair, Nucl. Phys. {\bf B336} (1990) 409}
\REF\BerLeCiii{D. Bernard and A. LeClair, Phys. Lett. {\bf B247} (1990) 309}
\REF\DeVKar{H.J. de Vega and M. Karowski, Nucl. Phys. {\bf B285} (1987) 619}
\REF\FenInt{P. Fendley and K. Intriligator, Nucl. Phys. {\bf B372} (1992) 533;
{\bf B380} (1992) 265}
\REF\FenSal{P. Fendley and H. Saleur, Nucl. Phys. {\bf B388} (1992) 609}
\REF\Zamoliii{Al.B. Zamolodchikov, Nucl. Phys. {\bf B358} (1991) 497}
\REF\Zamolod{A.B. Zamolodchikov, in Proc. Fields, strings
and quantum gravity, Beijing, China, 1989}
\REF\Schout{K. Schoutens, Nucl. Phys. {\bf B344} (1990) 665}
\REF\Ahn{C. Ahn,  Nucl. Phys. {\bf B354} (1991) 57}
\REF\ShaWit{R. Shankar and E. Witten, Phys. Rev. {\bf D17} (1978) 2134}
\REF\FanWu{C. Fan and F.Y. Wu, Phys. Rev. {\bf B2} (1970) 723}
\REF\Felderhof{B.U. Felderhof, Physica {\bf 65} (1973) 421; {\bf 66}
(1973) 279, 509}
\REF\Surth{B. Sutherland, J. Math. Phys. {\bf 11} (1970) 3183}
\REF\ChaKul{M. Chaichian and P. Kulish, Phys. Lett. {\bf B183} (1987) 169}
\REF\BabLan{O. Babelon and F. Langouche, Nucl. Phys. {\bf B290} [FS20]
(1987) 603}
\REF\Olsh{M.A. Olshanesky, Comm. Math. Phys. {\bf 88} (1983) 63}
\REF\FGS{S. Ferrara, L. Girardello, and S. Sciuto, Phys. Lett. {\bf B76}
(1978) 303}
\REF\SenMaj{S. Sengupta and P. Majumdar, Phys. Rev. {\bf D33}
(1986) 3138}
\REF\DiVFer{P. Di Vecchia and S. Ferrara, Nucl. Phys. {\bf B130}
(1977) 93}
\REF\KlaMel{T. Klassen and E. Melzer, Nucl. Phys. {\bf B338} (1990) 485;
{\bf B350} (1991) 635}
\REF\ChrMar{P. Christe and M.J. Martins, Mod. Phys. Lett. {\bf A5} (1990)
2189; M.J. Martins, Phys. Lett. {\bf 240B} (1990) 404}
\REF\Zamoliv{Al. B. Zamolodchikov, Phys. Lett. {\bf B253} (1991) 391}
\REF\IaM{H. Itoyama and P. Moxhay, Phys. Rev. Lett. {\bf 65} (1990)
2102}
\REF\AhnNam{C. Ahn and S. Nam, Phys. Lett. {\bf B271} (1991) 329}
\REF\CLT{S. Chung, E. Lyman, and H. Tye, Int. J. Mod. Phys. {\bf A7}
(1992) 3339}
\REF\ZamC{A.B. Zamolodchikov, JETP Lett. {\bf 43} (1986) 702;
Sov. J. Nucl. Phys. {\bf 46} (1987) 1090}
\REF\LudCar{A.W.W. Ludwig and J.L. Cardy, Nucl. Phys. {\bf B285}
(1987) 687}
\REF\Cardy{J.L. Cardy, Phys. Rev. Lett. {\bf 60} (9188) 2709}
\REF\Ahnnext{C. Ahn, in preparation}
\REF\BazStr{V.V. Bazhanov and Yu.G. Stroganov, Theor. Math. Phys.
{\bf 62} (1985) 253; {\bf 63} (1985) 519, 604}
\REF\CGLS{R. Cuerno, C. Gomez, E. Lopez-Manzanares, and G. Sierra,
Madrid preprint IMAFF-2/93 (1993).}

\chapter{Introduction}

For 2D integrable field theories $S$-matrices are purely elastic,
all incoming momenta are conserved and multi-particle scattering amplitudes
are factorized into a product of two-particle $S$-matrices.
These $S$-matrices, in turn, should satisfy Yang-Baxter equations which often
determine the $S$-matrices completely along with unitarity and crossing
symmetry [\ZamZamol].
The $S$-matrix provides essential tools to understand 2D field theories.
First of all, the $S$-matrix gives information on the UV behaviour of the
theory by relating the Casimir energy on the cylinder to
the central charge of the corresponding UV conformal field theory
(CFT) [\BPZ].
This program known as thermodynamic Bethe ansatz (TBA) [\Zamoli]
has provided
consisitency checks for many factorizable scattering theories either
with local lagrangians or without them such as perturbed CFTs [\Zamiii].

$S$-matrix plays an important role in off-shell physics as well.
It can be used to
determine off-shell quantities such as correlation
functions by computing the matrix
elements of an operator on the basis of the on-shell particles.
These objects known as form factors (FFs) may be computed exactly using only
the $S$-matrices and particle spectrum (bound states) as input
[\KarWei,\Smibook].
With exact FFs correlation functions are given by an infinite sum over
intermediate on-shell states.
This form factor approach has an advantage for the computation of
correlation functions of massive integrable models that the infinite
sum over all intermediate states converges very fast.
For many cases, upto two-point FFs give quite accurate results
on off-shell quantities [\YurZam--\CarMus].
Furthermore, the two-point FFs can be related to some
exact non-perturbative informations of the underlying theories,
such as the wave function renormalization [\KarWei,\DesDeV]
and the UV central
charges through the spectral representation of the $C$-theorem
[\Zamolii,\FLV,\FMS].
In this sense, without complete solutions of the FFs one can still
extract non-perturbative off-shell informations from the FFs.

While the TBA analysis or the FF computation can be relatively simple for
diagonal scattering theory, which has no mass degeneracy,
non-diagonal scattering theories entail much more complicacy.
By nondiagonal we mean theories with
different types of particles of the same mass for which
the scattering of two particles can occur in more than one channels.
Most of interesting 2D integrable field theories such as
the soliton scattering theories, theories with internal gauge symmetries,
and supersymmetric theories belong to this class.

For the non-diagonal theories, the equations for the TBA and FFs are
expressed in terms of monodromy and transfer matrices.
To solve the equations, one needs to diagonalize these  matrices.
It is remarkable that with some technical diffferences the same problem is
often met in the study of the lattice models [\Baxbook].
In the lattice model the Yang-Baxter equations are to be satisfied to
construct infinite number of conserved charges through the commuting
transfer matrices.
Partition functions and free energies are expressed in terms of the eigenvalues
of the transfer matrices.
Due to this common feature, it is often quite useful
to connect 2D field theories with lattice models.

There are two types of the models in the lattice and continuum which are
connected with each other.
The first one is so-called the vertex type; the states are assigned on the
lines
which form a lattice. For the square lattice, each vertex consists of four
lines and assigned a Boltzmann weight depending on the four states of the lines
[\Baxbook].
These lines correspond to the world lines of incoming and outgoing
particles in the scattering theories.
While some of these vertex models are associated with field theories with local
lagrangians, there remain many vertex-type lattice models still to
be related to 2D integrable field theories.

The second type is the interacting-round-face models [\ABF].
The Boltzmann weights are assigned
on each vertex on the square lattice, depending on the heights of four faces.
As a special case, if the heights are restricted, one obtains restricted
solid-on-solid (RSOS) type of models. These wide class of the lattice models
have been related to 2D CFTs.
Due to the conformal invariance, the corresponding lattice models are at
the criticality. Many exact results including correlation functions have
been obtained using the CFT techniques.
These identification can be continued in the off-critical region.
Without the conformal symmetry, the off-critical RSOS models are
associated with CFTs perturbed by relevent operators [\LeC--\ABL].
Again, $S$-matrices of the perturbed CFTs are given by the Boltzmann
weights of the RSOS models.

The best known example is the relation between the six vertex model and
the sine-Gordon (SG) model.
The SG model has soliton and antisoliton spectrum
and the $S$-matrix can be associated with
the $R$-matrix of the $\wh{{\rm sl}}_q(2)$, affine quantum group [\BerLeCii].
The Boltzmann weights of the six vertex model are the same as the $S$-matrix
elements  after identifying the up and down arrows assigned on each vertex line
with the soliton and antisoliton.
In addition, quantum group reduction of the SG model corresponds
to the RSOS lattice model obtained from the six  vertex model.
The TBA analysis of these models have been done by diagonalizing the
inhomogeneous transfer matrices of the six vertex [\DeVKar,\FenInt,\FenSal]
and RSOS models [\Zamoliii].

The complete FFs of the SG model have been obtained by F. Smirnov using
quantum inverse scattering methods, providing only known example with the
complete FFs for nondiagonal theories.
Based on this information, Smirnov found axioms for the FFs to satisfy
[\Smibook].
Therefore, the problem to find complete FFs is reduced to solve these axioms
for a given theory.
However, solving these axiomatic equations completely is very difficult
even for diagoanal scattering theories except a few simplest ones such as
Ising, Yang-Lee, and sinh-Gordon models [\Zamolii,\YurZam,\FMS,\BabBer].
The problem becomes much more complicated for the nondiagonal cases.
As an initial step to the problem, we will concentrate on two-point FFs.
Two-point FFs can be determined relatively easily
by diagonalizing $S$-matrix and evaluating the FFs using the Watson
equations [\KarWei]. For the supersymmetric theories, details can be further
simplified due to the SUSY relations between the FFs.
As stressed before, the two-point FFs have many useful informations on
the underlying theories.

In this paper, we want to apply these frameworks to the $N=1$
supersymmetric (SUSY) theories.
The $S$-matrices of many SUSY models have been otained. These
$S$-matrices have the following factorized form [\ABL,\Schout]:
$$S(\t)=S_{\rm S}(\t)\otimes S_{0}(\t),\eqn\eqi$$
where the first factor $S_{\rm S}$ carries the SUSY indices and commutes
with the SUSY charges while the second one $S_{0}$ is the $S$-matrices of
the models without the SUSY.
So far, several SUSY integrable field theories and perturbed super CFTs are
solved and their $S$-matrices are derived.
Interesting aspect of the SUSY models is that these $S$-matrices commuting with
SUSY charges are identified with Boltzmann weights of some lattice models.

For example, for the $N=1$ SUSY CFTs perturbed by the least relevent
operator, $S_{\rm S}$, which commutes with SUSY charges with central extension
due to the topological charges, is related to the RSOS weights corresponding to
the tricritical Ising model [\Zamolod].
For the $N=2$ SUSY models, the first factor is identified with the Boltzmann
weights of the six vertex model [\BerLeCiii,\FenInt].
These relations with lattice models are important not only for the lattice-
field theory correspondence but for actual solutions of the models.

$N=1$ SUSY sine-Gordon (SSG) model has been solved in an unconventional way.
Its soliton $S$-matrix has been derived from the results on
the perturbed super CFTs by the least relevent operator [\ABL].
The SUSY part of the SSG soliton $S$-matrix is given by the RSOS
tricritical Ising model $S$-matrix while $S_0$ is ordinary sine-Gordon
$S$-matrix.
The $S$-matrices of the SSG bound states (breathers) have been derived
from multi-soliton scattering amplitudes [\Ahn].
In particular, since the lightest bound states are forming a supermultiplet of
the fundamental fields appearing in the SSG lagrangian, the lightest
breather $S$-matrix of the SSG model can be analytically continued to get
the $S$-matrix of supersymmetric sinh-Gordon (SShG) model.
This $S$-matrix is identical to the one derived first by Shankar and Witten
by explicitly requiring the commutativity with SUSY charges [\ShaWit].
Besides, the SSG model with only the lightest breather in the spectrum can
be understood as perturbed super CFTs, the SUSY Yang-Lee (SYL) model
[\Schout,\Ahn];
the simplest nonunitary super CFT perturbed by the least relevent operator.
This model includes only one supermultiplet of on-shell states and
the $S$-matrix is identical with that of the SShG model.
This $S$-matrix is our starting point.

These models with $N=1$ SUSY without a central extension
will be identified with the general eight vertex models with an external field.
If the Boltzmann weights of the general  eight  vertex model
satisfy a `free fermion' condition, the model
is exactly solvable and the free energy was derived first from dimer method
[\FanWu] and later diagonalizing the transfer matrix [\Felderhof].
Also, this model has been identified with the general $XY$-spin chain
model with an magnetic field [\Surth].
This relation with the lattice model
will be very useful in our derivation of TBA equations for the SShG model.
It turns out that the SShG model is at the critical point of the $XY$-spin
chain model.

We organize this paper in the following order.
In next section, we write down the lagrangian of the SShG and SSG models
and derive the energy-momentum tensor supermultiplet and their relations
under the SUSY transformation.
Also we present the $S$-matrices of the models.
In sect.3, we review the basic formulae of TBA analysis which will be
used in the next section where explicit derivations are explained.
In sect.5, we construct the FFs of the SShG model using the Watson equations
and SUSY relations of the energy-momentum tensor.
With exact two-point FFs, we derive the UV central charge
of the model using the spectral representation of the $C$-theorem.

\endpage
\chapter{$N=1$ SUSY Integrable Model and Factorizable $S$-Matrix}

We present the energy-momentum tensor supermultiplet
of the $N=1$ SSG and SShG model and the $S$-matrix of the theories.

\section{Lagrangian and Energy-Momentum Tensor}

We start with a langrangian of a general $N=1$ SUSY
$${\cal L}(\Phi)={1\over{4}}{\ol D}\Phi D\Phi
+iW(\Phi)\bigg\vert_{\t_1\t_2},\eqn\eqL$$
with a scalar superfield $\Phi$
$$\Phi(x,\t)=\p+i{\ol\t}\psi+i\half{\ol\t}\t F,\eqn\eqP$$
and $D$ and ${\ol D}$, the covariant derivatives
$$D_{\a}={\pd\over{\pd{\ol\t}^{\a}}}+i(\g^{\m}\t)_{\a}\pd_{\m}.\eqn\eqD$$
The Grassman variable $\t$ is a Majorana spinor.
\foot{Dirac matrices are $\g^{0}=\pmatrix{0&1\cr -1&0\cr}\quad
\g^{1}=\pmatrix{0&1\cr 1&0\cr}$.}
In terms of the component fields, one gets
$${\cal L}(\Phi)=\half(\pd_{\m}\p)^2+{i\over{2}}{\ol\psi}\left[
\sl{\pd}+W''(\p)\right]\psi+\half\left[W'(\p)\right]^2 .\eqn\eqLag$$
The SSG model is a particular case of Eq.\eqLag\ with the superpotential
$$W(\P)={m\over{\b^2}}\cos(\b\P).\eqn\eqSP$$
The SShG  model is the same superpotential with the purely imaginary coupling
constant $\b=i\wh{\b}$. The $N=1$ SUSY algebra is generated by the conserved
charges $Q_1$ and $Q_2$
$$Q_1^2=P_{+},\quad Q_2^2=P_{-},\quad{\rm and}\quad \{Q_1,Q_2\}=0,\eqn\eqSUSY$$
with the light-cone momenta defined as $P_{\pm}=E\pm P$.
These charges act on the component fields by
$$\eqalign{
Q_1\p&=i\psi_1,\qquad Q_1\psi_1=\ \pd_{+}\p,\qquad Q_1\psi_2=\ F,\cr
Q_2\p&=i\psi_2,\qquad Q_2\psi_2=-\pd_{-}\p,\qquad Q_2\psi_1=-F,\cr}\eqn\susyi$$
with $F=-W'(\p)$.

Integrability of the SSG and SShG models is estabilished because they are
equivatent to Toda theory based
on the twisted super affine Lie algebra $C^{(2)}(2)$
[\ChaKul--\Olsh]. The equations of motion of the SSG theory can be rewritten
as super zero-curvature conditions.
An infinite number of conserved charges at the classical level were derived
[\FGS] and checked to be preserved at the lowest order
quantum corrections [\SenMaj].

The energy-momentum tensor supermultiplet can be expressed by [\DiVFer],
$$J_{\a\m}=\left[\left(\sl{\pd}\Phi-W'(\Phi)\right)\g_{\m}D
\Phi\right]_{\a},\eqn\eqEM$$
or in light-cone coordinates
$$J_{+}=\pmatrix{D_1\Phi\pd_{+}\Phi\cr
    -W'(\Phi) \Phi D_1\Phi\cr},\quad
    J_{-}=\pmatrix{ -W'(\Phi)\Phi D_2\Phi\cr
     D_2\Phi\pd_{-}\Phi,\cr}\eqn\eqJJ$$
with $x_{\pm}=\half(x_1\pm x_0)$ and $\pd_{\pm}=\pd_1\pm\pd_0$.
In terms of the component currents,
$$J_{\pm}=\pmatrix{\Psi_{1\pm}\cr
        \Psi_{2\pm}\cr}+2i\pmatrix{
        \t_2 T_{+\pm}\cr  \t_1 T_{-\pm} \cr}
        +i\t_1\t_2\pmatrix{\chi_{1\pm}\cr \chi_{2\pm}\cr},\eqn\eqJ$$
one gets the energy-momentum tensor of the SSG model
$$\eqalign{
T_{++}&=\half\left[(\pd_{+}\p)^2+i\psi_1\pd_{+}\psi_1\right],\quad
T_{--}=\half\left[(\pd_{-}\p)^2-i\psi_2\pd_{-}\psi_2\right],\cr
T_{+-}&=T_{-+}=\half {m^2\over{\b^2}}\sin^2\b\p-{im\over{4}}\
\ol{\psi}\psi\ \cos\b\p,\cr}\eqn\eqTssg$$
and its  superpartner
$$\eqalign{
\Psi_{1+}&=i\psi_1\pd_{+}\p\qquad \Psi_{2+}=-i{m\over{\b}}\psi_1\sin\b\p\cr
\Psi_{2-}&=i\psi_2\pd_{-}\p\qquad \Psi_{1-}=-i{m\over{\b}}\psi_2\sin\b\p.\cr}
\eqn\eqsTssg$$
Including an appropriate normalization factor of $4\pi$,
we define the following notation for the SUSY energy-momentum tensor:
$$T=4\pi T_{++},\quad\ol{T}=4\pi T_{--},\quad\T=4\pi T_{+-},\eqn\eqTT$$
and their SUSY partners,
$$T_{\rm F}=4\pi \Psi_{1+},\quad\ol{T}_{\rm F}=4\pi \Psi_{2-},\quad\T_{\rm F}
=4\pi\Psi_{1-},\quad\ol{\T}_{\rm F}=4\pi\Psi_{2+}.\eqn\eqsTT$$
They are  related to each other by the SUSY transformation
$$\eqalign{
&Q_1T_{\rm F}=-2i T\quad  Q_1 T=-\half\pd_{+}T_{\rm F}\quad Q_1\T_{\rm F}
=-2i\T \quad Q_1\T=-\half\pd_{+}\T_{\rm F}\cr
&Q_2\ol{T}_{\rm F}=2i\ol{T}\quad Q_2 \ol{T}=-\half\pd_{-}\ol{T}_{\rm F}\quad
Q_2\ol{\T}_{\rm F}=2i\T\quad Q_2\T=-\half\pd_{-}\ol{\T}_{\rm F}.\cr}
\eqn\eqTTsusy$$

\section{On-Shell Particle States and $S$-Matrix}

If the coupling constant of the SSG model in
Eq.\eqSP\ becomes pure imaginary, we have a simplest $N=1$ SUSY field theory,
namely the SShG model.
Since the potential is not periodic, the soliton spectrum does not exist
any more and the spectrum consists of only
the fundamental particles appearing in the lagrangian, one scalar and fermion
supermultiplet.
We will denote on-shell states of these particles
by $\ket{b(\t)}$ and $\ket{f(\t)}$ with a
rapidity $\t$ which is related to the momentum by $E=m\cosh\t$ and
$P=m\sinh\t$.

The SUSY charges defined in Eq.\eqSUSY\ can act on on-shell states as
(See sect.5.3)
$$\eqalign{&Q_1\ket{f(\t)}=\sqrt{m}e^{\t/2}\ket{b(\t)},\qquad
Q_1\ket{b(\t)}=\sqrt{m} e^{\t/2}\ket{f(\t)},\cr
&Q_2\ket{f(\t)}=-i\sqrt{m} e^{-\t/2}\ket{b(\t)},\qquad
Q_2\ket{b(\t)}=i\sqrt{m} e^{-\t/2}\ket{f(\t)}.\cr}\eqn\onshellsusy$$
It is easy to see that this satisfies $N=1$ SUSY algebra, Eq.\eqSUSY.
Action of SUSY charges on multiparticle on-shell states can be easily
worked out using this and the anticommutivity of $Q_{\a}$ and the fermion.

Exact $S$-matrix of the SShG model was derived using the Yang-Baxter
equation, unitarity and crossing symmetry along with the commutativity of
the SUSY charges and the $S$-matrix [\ShaWit].
In the basis of two-particle on-shell states in the order of
$\ket{b_1 b_2},\ket{f_1 f_2},\ket{b_1 f_2},\ket{f_1 b_2}$,
\foot{
We use a short notation $\ket{b_1 b_2}=\ket{b(\t_1)b(\t_2)}$ etc.}
the $S$-matrix has been obtained to be ($\t=\t_1-\t_2$):
$$S(\t)=Y(\t)\pmatrix{1+{2i\sin\a\pi\over{\sinh\t}}&
{i\sin\a\pi\over{\cosh{\t\over{2}}}}&0&0\cr
{i\sin\a\pi\over{\cosh{\t\over{2}}}}&1-{2i\sin\a\pi\over{\sinh\t}}&0&0\cr
0&0&1&{i\sin\a\pi\over{\sinh{\t\over{2}}}}\cr
0&0&{i\sin\a\pi\over{\sinh{\t\over{2}}}}&1\cr},\eqn\Sshawit$$
with an arbitrary constant $\a$ which will be related to the coupling constant
$\b$ of the SSG model in a moment.
The prefactor $Y(\t)$ is needed to make the $S$ matrix unitary and crossing
symmetric.
The following integral form will be useful later:
$$Y(\t)={\sinh{\t\over{2}}\over{\sinh{\t\over{2}}+i\sin(|\a|\pi)}}\exp
-\int_{0}^{\infty}{dt\over{t}}{\sinh(|\a| t)
\sinh((1-|\a|)t)\over{\cosh^2{t\over{2}}\cosh t}}
\sinh{\t t\over{\pi i}}.\eqn\eqYint$$
With $Y(\t)=Y(i\pi-\t)$ and a factor of $i$ arising in the crossing relation
for
$bb\to ff$ channel,
the $S$-matrix of Eq.\Sshawit\ is crossing symmetric.

To determine the constant $\a$ we should refer to another derivation of the
SSG breather $S$-matrix. Using the SSG soliton $S$-matrix,
one can compute four soliton scattering amplitudes.
By taking bound state poles of the incoming and outgoing soliton-antisoliton
pairs one can derive the $S$-matrices of the SSG breathers [\Ahn].
In particular the $S$-matrix of the lightest breathers is the $S$-matrix of
the fundamental particles of the SSG and SShG models:
$$\eqalign{S(\t)&=Y(\t)\cd{\cal R}(\t)\cd S_{0}(\t)\cr
{\rm where}\qquad S_{0}(\t)&={\sinh\t+i\sin(2\a\pi)\over{
\sinh\t-i\sin(2\a\pi)}},\cr
{\cal R}(\t)&=\pmatrix{1+{2i\sin\a\pi\over{\sinh\t}}&
{\sin\a\pi\over{\cosh{\t\over{2}}}}&0&0\cr
{\sin\a\pi\over{\cosh{\t\over{2}}}}&-1+{2i\sin\a\pi\over{\sinh\t}}&0&0\cr
0&0&1&{i\sin\a\pi\over{\sinh{\t\over{2}}}}\cr
0&0&{i\sin\a\pi\over{\sinh{\t\over{2}}}}&1\cr}.\cr}\eqn\Sahn$$
The factor $S_0$ is the lightest breather $S$-matrix of the SG model.
The constant $\a$ in Eq.\Sahn\ is given by the coupling constant of the SSG
model [\Ahn]
$$\a={\g\over{16\pi}}={\b^2/4\pi\over{1-\b^2/4\pi}}.\eqn\ssgcoup$$
For the SShG model with $\b=i\wh{\b}$ ($\wh{\b}$ real), this constant
reduces to
$$\a=-{\wh{\b}^2/4\pi\over{1+\wh{\b}^2/4\pi}},\eqn\sshgcoup$$
and $-\half<\a<0$.

Two $S$-matrices, Eqs.\Sshawit\ and \Sahn\  are equivalent.
The sign difference in the $ff\to ff$ channel is explained because
all particles are considered as bosons in Eq.\Sahn\ by including
the exchange factor $-1$ arising in $ff\to ff$
in the $S$-matrix element.
In this convention, the crossing relation is satisfied without any extra factor
because all particles are bosonic.
Besides,
for the SShG model with $\a<0$, the $S_0$ has no pole in the physical strip.
Therefore, $S_0$ is nothing but a CDD factor and can be removed under
the minimality assumption.
For the SSG model, however, with a coupling in  $0<\a<\half$ ($\b^2<4\pi/3$)
the $S_0$ does have a bound state pole corresponding to the second breather.

For a complete description of the SSG model, one should include all
the $S$-matrices of the solitons and breathers as was done in [\Ahn].
Depending on the values of the coupling constant of the SSG model, the
spectrum of the bound states changes.
In particular, if the coupling constant is in the range of
$\half<{\g\over{8\pi}}<1$, only the lightest
bound states can exist along with the soliton and antisoliton in the spectrum.
If the solitons are truncated from the theory keeping only the lightest
bound states, the scattering theory becomes perturbed CFT by the least relevent
operator. The UV CFT is the SUSY extension of the Yang-Lee model
[\Schout,\Ahn].
The $S$-matrix is given by Eq.\Sahn.
\endpage

\chapter{Thermodynamic Bethe Ansatz}

\section{Diagonal TBA}

The TBA computes the Casimir energy for a theory on a circle of length $R$
with $S$-matrices and particle spectrum as input data [\KlaMel--\AhnNam].
With a temperature $T=1/R$ the configuration of minimizing free energy gives
the ground state energy of the system, which is again
related to the central charge of the underlying UV CFT by
$$E(R)\approx -{\pi\over{6R}}\left(C-12\D_{\rm min}-12\ol{\D}_{\rm min}\right),
\eqn\Ccharge$$
as $R\to 0$ (or $T\to\infty$).
$\D_{\rm min}$ (and $\ol{\D}_{\rm min}$) stands for the lowest conformal
dimension allowed by the conformal theories.
For unitary theories $\D_{\rm min}$ is zero for the identity operator while
it is in general negative for nonunitary theories.

Consider $N+1$ particles in a box of length $L$ with periodic boundary
condition (PBC).
If we move the $(N+1)$-st particle of mass $m_a$ and rapidity $\t_k$
exchanging with all the other particles and come back to the original
configuration, we get the following PBC equation:
$$e^{im_aL\sinh\t_k} \prod_{i=1}^{N} S_{a a_i}(\t_k-\t_i)=1,\eqn\eqPBC$$
where the index $a_i$ specifies species of the $i$-th particle.
For the diagonal scattering theories, the product of $S$-matrices are just
$C$-number and one can take logarithms on both sides to get
$$m_a L\sinh\t_k + \sum_{i=1}^{N} {1\over{i}}\ln S_{a a_i}(\t_k-\t_i)=
2\pi n_k,\eqn\tbai$$
with an arbitrary integer $n_k$.
These transcedental equations give solutions for $\{\t_i\}$ for a given
set of arbitrary integers $\{n_k\}$.
Therefore, considering all possible integers, and
as $N,\ L\to\infty$ the solutions form a band structure and one can
introduce a density of the rapidity states, $\r_a(\t)$, defined by the number
of
allowed rapidity states between $\t$ and $\t+d\t$ divided by $d\t$.
Any $N$ rapidities of these states can be a solution of the PBC.

Therefore, Eq.\tbai\ can be expressed by
$$2\pi\r_a(\t)=m_a L\cosh\t+\sum_{b}\int d\t' \r^{1}_b(\t')\p_{ab}(\t-\t'),
\eqn\tbaii$$
where the $\r_a^{1}(\t)$ is the density of rapidity states which are actually
occupied by the on-shell particles and
$$\p_{ab}(\t)={1\over{i}}{\pd\over{\pd\t}}\ln S_{ab}(\t).$$
Introducing `psuedo-energy' $\ep_a$ defined by
$${\r^{1}_a(\t)\over{\r_a(\t)}}={e^{-\ep_a(\t)}\over{1+e^{-\ep_a(\t)}}},
\eqn\tbaiii$$
one can express the ground state energy by
$$E(R)=-\sum_a m_a\int_{-\infty}^{\infty} {d\t\over{2\pi}}
\cosh\t\ln\left(1+e^{-\ep_a(\t)}\right), \eqn\tbaiv$$
where $\ep_a$ is determined by the minimizing condition of the free energy:
$$\ep_a(\t)=m_a R\cosh\t-\sum_b\int_{-\infty}^{\infty}
{d\t'\over{2\pi}}\p_{ab}(\t-\t')\ln\left(1+e^{-\ep_b(\t')}\right).\eqn\tbav$$
This TBA equation can be solved easily for the UV ($R\to 0$) and IR
($R\to\infty$) cases because the equations can be effectively described by
simple algebraic equations.

\section{Nondiagonal TBA}

For the nondiagonal theories, the product of $S$-matrices in
Eq.\eqPBC\ is the monodromy matrix and the PBC equation can be expressed as
$$\eqalign{
&e^{im_a L\sinh\t}{\cal T}_{aa}(\t|\t_1,\ldots,\t_N)=1,\qquad{\rm where}\cr
&{\cal T}_{ab}(\t|\t_1,\ldots,\t_N)_{\{a_i\}}^{\{a'_i\}}=\sum_{\{\a_i\}}
S_{a a_1}^{\a_2 a'_1}(\t-\t_1) S_{\a_2 a_2}^{\a_3 a'_2}(\t-\t_2)
\cdots S_{\a_N a_N}^{b a'_N}(\t-\t_N).\cr}\eqn\eqPBCi$$
If we add these equations for the index $a$,
we can express it in terms of the transfer matrix
$$e^{imL\sinh\t} T(\t|\t_1,\ldots,\t_N)=N_c,\eqn\tbavi$$
where the integer $N_c$ is the number of colors and the transfer matrix
$T\equiv\sum{\cal T}_{aa}$ acts on $V^{\otimes N}$.
Precisely speaking, this is `inhomogeneous' transfer matrix because it depends
on each rapidity of in-coming particle states.

To derive the TBA equations, one must diagonalize the transfer matrix
which is quite a difficult task due to the size of the matrix and
the inhomogeneity.
In the lattice models, many pretty ideas have been invented for the
purpose [\Baxbook].
Although these methods are, in principle, applicable only to homogeneous cases,
some minor corrections make it possible to apply it to the inhomogeneous ones.
Among these, two methods have been successfully applied to derive TBA
equations.
The first one is using the inversion relation of the transfer matrix.
TBA equation for the RSOS model corresponding to $N=1$ SUSY soliton
scattering theory has been obtained in this way [\Zamolii].
Due to the periodic property of the transfer matrix and its inverted matrix,
the eigenvalues are completely fixed by the location of `zeroes'.
These zeroes satisfy constraint equations in terms of the rapidities $\t_i$'s.

Another useful approach is based on the algebraic Bethe ansatz method.
In this method, one construct the eigenstates in terms of the monodromy matrix
element acting on the vacuum.
The eigenvalues can be directly obtained with an additional contraint which
makes the eigenstate ansatz to be actual eigenstate.
For the example, this method can be used for the SG model and $N=2$ SUSY
theories [\DeVKar--\FenSal].

In general, the eigenvalues of the transfer matrix have the form like
$$\L(\t|\t_1,\ldots,\t_N)=\prod_{k=1}^{N} f(\t-x_k),\eqn\tbavii$$
with the zeroes $x_k$ of a function $f(x)$ ($f(0)=0$) which satisfy
$$\prod_{i=1}^{N} g(x_k-\t_i)={\rm Const}.\eqn\tbaviii$$
By taking an imaginary part of the logarithms of both sides of Eq.\eqPBCi\
and introducing pseudo-energies both for the real particle states and for
the zeroes, one can find the TBA equations which are very similar with
those of the diagonal TBA.
Only difference is there is no mass term (`driving term') in the TBA
equation for the density of the zeroes.
This absence of the driving term makes a big difference in the analysis of the
TBA equations.

\section{Casimir Energy and  UV Central Charge}

The TBA equations in the UV limit ($R\to 0$) can be easily solved
because the pseudo-energies become independent of the rapidity around $\t=0$.
This plateaux extends upto $\t\sim -\ln (mR)$.
Therefore, the pseudo-energies become practically constant in the limit
and the TBA equations can be reduced to the mere algebraic equations like
$$x_a=\prod_b (1+x_b)^{N_{ab}},\eqn\tbaix$$
with $x_a=\exp(-\ep_a(0))$ and
$$N_{ab}={1\over{2 \pi}}\left[\p_{ab}(\infty)-\p_{ab}(-\infty)\right].
\eqn\tbax$$
If all mass terms are non-zero, the $\ep_a(\infty)$ diverges like
$\ep_a(\t\to\infty)\sim m_a R\cosh\t$ and $y_a=\exp(-\ep_a(\infty))$ vanishes.
If some of the driving terms are zero as is the case for nondiagonal TBA,
$y_{a'}$ satisfies
$$y_{a'}=\prod_{b'} (1+y_{b'})^{N_{a'b'}},\eqn\tbaxi$$
where $a'$ denotes species with vanishing driving term.
The ground state energy can be expressed compactly with these variables
$x_a$ and $y_a$ by
$$E(mR)\sim -{1\over{\pi R}}\sum_{a}\left[{\cal L}
\left({x_a\over{1+x_a}}\right)-{\cal  L}\left({y_a\over{1+y_a}}\right)\right]
\eqn\tbaxii$$
where  ${\cal L}(x)$ is the Rogers dilogarithmic function
$${\cal L}(x)=-\half\int^{x}_{0}dt\left[{\ln(1-t)\over t} +
{\ln t\over{(1-t)}}\right].\eqn\tbaxiii$$
{}From Eq.\Ccharge\ the central charge of the UV CFT is given
by
$$C-12(\D_{\rm min}+\ol{\D}_{\rm min})={6\over{\pi^2}}
\sum_{a}\left[{\cal L}
\left({x_a\over{1+x_a}}\right)-{\cal  L}\left({y_a\over{1+y_a}}\right)\right].
\eqn\tbaxiv$$
\endpage

\chapter{TBA for the $N=1$ SUSY models}

In this section we diagonalize the inhomogeous transfer matrix associated
with the SShG $S$-matrix.
The essential observation is that Eq.\Sahn\ satisfies so-called `free fermion'
condition of the general eight vertex model with an external field.
We will derive the TBA equation based on the inversion relation for the
transfer matrix.
We apply this TBA equation to both the SShG model and a SYL model
perturbed by the least relevent operator and derive correct UV central charges.

\section{Free Fermion Models}

After the celebrating  solution of the symmetric eight vertex model by Baxter,
 Fan and Wu obtained an exact expression of the free energy for
the general eight vertex model with an external field if the Boltzmann weights
satisfy some additional contraint, named the free fermion condition [\FanWu].
They called this model `Free Fermion' model (FFM) although the name is
slightly misleading.
The model turned out to be highly non-trivial and interacting.

We start with the Boltzman weights of the general eight vertex model:
$$R=\pmatrix{a_{+}&0&0&d\cr 0&b_{+}&c&0\cr 0&c&b_{-}&0\cr
             d&0&0&a_{-}\cr},\eqn\ffmi$$
for the following vertex configurations:
\def\la{\leftarrow}
\def\ra{\rightarrow}
\def\ua{\uparrow}
\def\da{\downarrow}
$$ \vbox{\settabs 8 \columns
\+\hbox{\hbox{\kern 3pt$\ra$}\hbox{\kern-3pt${\ua\atop{\ua}}$}
\hbox{\kern-6pt$\ra$}    }
&\hbox{\hbox{\kern 3pt$\la$}\hbox{\kern-3pt${\da\atop{\da}}$}
\hbox{\kern-6pt$\la$}    }
&\hbox{\hbox{\kern 3pt$\ra$}\hbox{\kern-3pt${\da\atop{\da}}$}
\hbox{\kern-6pt$\ra$}    }
&\hbox{\hbox{\kern 3pt$\la$}\hbox{\kern-3pt${\ua\atop{\ua}}$}
\hbox{\kern-6pt$\la$}    }
&\hbox{\hbox{\kern 3pt$\ra$}\hbox{\kern-3pt${\ua\atop{\da}}$}
\hbox{\kern-6pt$\la$}    }
&\hbox{\hbox{\kern 3pt$\la$}\hbox{\kern-3pt${\da\atop{\ua}}$}
\hbox{\kern-6pt$\ra$}    }
&\hbox{\hbox{\kern 3pt$\ra$}\hbox{\kern-3pt${\da\atop{\ua}}$}
\hbox{\kern-6pt$\la$}    }
&\hbox{\hbox{\kern 3pt$\la$}\hbox{\kern-3pt${\ua\atop{\da}}$}
\hbox{\kern-6pt$\ra$}    }\cr
\+\quad$a_{+}$&\quad$a_{-}$&\ \quad$b_{+}$&\ \quad$b_{-}$&
\ \quad$c$&\ \quad$c$&\ \quad$d$&\ \quad$d$&\cr}\eqn\ffmBWs$$

If $R(\t)$ satisfies the Yang-Baxter equation and the free fermion condition
$$a_{+}a_{-}+b_{+} b_{-}=c^2 + d^2,\eqn\ffmii$$
and if the following combination of the Boltzman weights are independent of
the  rapidity
$$\G={2cd\over{a_{+}b_{-}+a_{-}b_{+}}},\qquad
h={a_{-}^2+b_{+}^2-a_{+}^2-b_{-}^2\over{2(a_{+}b_{-}+a_{-}b_{+})}},\eqn\ffmiii$$
the transfer matrix $T$ commutues; $[T(u),T(v)]=0$.
Due to this commutativity, there exist infinite number of conserved charges
including a Hamiltonian of the corresponding one-dimensional spin-chain model.
This Hamiltonian has been identified with that of the $XY$-model with a
magnetic
field,
$${\cal H}_{XY}=-J\sum_{j=1}^{N}[\s^{+}_{j}\s^{-}_{j+1}+\s^{-}_{j}\s^{+}_{j+1}
 +\G(\s^{+}_{j}\s^{+}_{j+1}+\s^{-}_{j}\s^{-}_{j+1})-h\s^{z}_{j}],\eqn\ffmiv$$
where $\s^{\pm}=\half(\s^{x}\pm i\s^{y})$ with a conventional Pauli $\s^{i}$
matrices.

To identify the FFM with $N=1$ SShG model,
we rewrite the $S$-matrix of the SShG model, Eq.\Sahn, by rearranging the
two-particle basis.
In the order of $\ket{bb},\ket{bf},\ket{fb},\ket{ff}$,
the $R$-matrix of Eq.\Sahn\ becomes the general form of the FFM with
$$a_{\pm}=\pm 1+{2i\sin\a\pi\over{\sinh\t}},\quad b_{\pm}=1,\quad
c={i\sin\a\pi\over{\sinh{\t\over{2}}}},\quad
d={\sin\a\pi\over{\cosh{\t\over{2}}}},\eqn\ffmv$$
if we identify $\uparrow$ and $\rightarrow$ with $\ket{b}$  and
$\downarrow$ and $\leftarrow$ with $\ket{f}$.
\foot{The Boltzman weights in Eq.\ffmBWs\ become the $S$-matrix element if
we adopt a convention that time flows from bottom-left to top-right
($\nearrow$).}

It is an easy exercise to check these weights satisfy the free fermion
condition Eq.\ffmii.
Also, the constants $\G$ and $h$ becomes
$$\G=\sin\a\pi,\qquad h=-1.\eqn\ffmvi$$
Since $h=-1$ is a critical point of the $XY$-model, the SShG model corresponds
to the critical point of the general eight vertex model with free fermion
condition and with vanishing modulus.

\section{Inversion Relation}

Critical step to derive TBA equations for the SShG model is to find an
inversion
relation for the transfer matrix.
We derive the following inversion relation in the Appendix A:
$$\eqalign{&\ T(u|\t_1,\ldots,\t_N)\ T(u+i\pi|\t_1,\ldots,\t_N)=(-1)^{N}
\times\cr
&\left[\prod_{i=1}^{N} M_{+}(u-\t_i)+\prod_{i=1}^{N} M_{-}(u-\t_i)+{F}
\left(\prod_{i=1}^{N}F_{+}(u-\t_i)+\prod_{i=1}^{N}F_{-}(u-\t_i)\right)\right],
\cr}\eqn\ffmvii$$
where  the fermion index operator $F$ is either $+1$ for the bosonic state
or $-1$ for the fermionic one.

The functions appearing in Eq.\ffmvii\ are expressed in terms of the Boltzmann
weights as follows:
$$\eqalign{&M_{+}=a_{+} a_{-}-d^2,\qquad M_{-}=a_{+} a_{-}-c^2,\cr
&F_{+}=\sinh^2\p\ a_{+}b_{+}+\cosh^2\p\ a_{-}b_{-}-2\sinh\p\cosh\p\ cd,\cr
&F_{-}=-\cosh^2\p\ a_{+}b_{+}-\sinh^2\p\ a_{-}b_{-}+2\sinh\p\cosh\p\ cd,\cr}
\eqn\ffmviii$$
and
$$\tanh(2\p)={2cd\over{a_{+}b_{+}+a_{-}b_{-}}}=\sin\a\pi.\eqn\ffmext$$
Using  Eq.\ffmv\ one can find
$$\eqalign{&M_{+}=-{\sinh\left({\t\over{2}}+i\a\pi\right)
\over{\sinh{\t\over{2}} }}{\sinh\left({\t\over{2}}-i\a\pi\right)
\over{\sinh{\t\over{2}} }},\
M_{-}=-{\cosh\left({\t\over{2}}+i\a\pi\right)\over{\cosh{\t\over{2}} }}
{\cosh\left({\t\over{2}}-i\a\pi\right)\over{\cosh{\t\over{2}} }},\cr
&F_{+}=-{\cosh\left({\t\over{2}}+i\a\pi\right)\over{\cosh{\t\over{2}}}}
{\sinh\left({\t\over{2}}-i\a\pi\right)\over{\sinh{\t\over{2}} }},
\  F_{-}=-{\sinh\left({\t\over{2}}+i\a\pi\right)\over{\sinh{\t\over{2}} }}
{\cosh\left({\t\over{2}}-i\a\pi\right)\over{\cosh{\t\over{2}}}}.\cr}\eqn\ffmix$$

{}From these expressions one can notice that under the change $u\to u+i\pi$
$$M_{\pm}\to M_{\mp}\qquad{\rm and}\qquad F_{\pm}\to F_{\mp},$$
therefore, $T(u)T(u+\pi i)=T(u+\pi i)T(u+2\pi i)$. This means
$$T(u|\t_1,\ldots,\t_N)= T(u+2\pi i|\t_1,\ldots,\t_N).\eqn\ffmx$$
These matrix relations can be easily transformed to equations of the
eigenvalues of the transfer matrices;
$\L(u|\t_1,\ldots,\t_N)$ is $2\pi i$ symmetric function,
$$\L(u|\t_1,\ldots,\t_N)=\L(u+2\pi i|\t_1,\ldots,\t_N),\eqn\ffmxi$$
and the inversion relation is nicely factorized,
$$\eqalign{&\L(u|\t_1,\ldots,\t_N)\L(u+\pi i|\t_1,\ldots,\t_N)\cr
&=\left[\prod_{i=1}^{N}{\cosh\left({u-\t_i\over{2}}+i|\a|\pi\right)\over{
\cosh\left({u-\t_i\over{2}}\right)}}+ F\prod_{i=1}^{N}
{\sinh\left({u-\t_i\over{2}}+i|\a|\pi\right)\over{\sinh\left({u-\t_i\over{2}}
\right)}}\right]\cr
&\times \left[\prod_{i=1}^{N}{\cosh\left({u-\t_i\over{2}}-i|\a|\pi\right)
\over{\cosh\left({u-\t_i\over{2}}\right)}}+ F\prod_{i=1}^{N}
{\sinh\left({u-\t_i\over{2}}-i|\a|\pi\right)\over{
\sinh\left({u-\t_i\over{2}}\right)}}\right].\cr}\eqn\ffmxii$$

\section{Eigenvalues}

Since $\L(u)$ is a $2\pi i$-periodic function with poles at
$u=\t_k$ and $u=\t_k+i\pi$, it can be completely fixed by the location of
zeroes on the strip in the complex plain of $-i\pi<{\rm Im}[u]\le i\pi$ and
$-\infty<{\rm Re}[u]<\infty$.
Also from the fact that $\L(u)$ becomes a constant as $u\to\infty$, we can
find that
$$\L(u|\t_1,\ldots,\t_N)={\rm Const.} \prod_{k=1}^{N}\left[{\sinh
\left({u-z^{+}_k\over{2}}\right)\over{\sinh\left({u-\t_k\over{2}}\right)}}
{\sinh\left({u-z^{-}_k\over{2}}\right)\over{\cosh
\left({u-\t_k\over{2}}\right)}}\right],\eqn\ffmxiii$$
where the $2N$ zeroes $\{z^{+}_k\}$ and $\{z^{-}_k\}$ located on the strip
will be determined as functions of $\t_i$'s.

We defined the zeroes in the way that
$z_k^{+}$ and $z_k^{-}$ come from the first and second factors
of the RHS of Eq.\ffmxii, respectively.
Therefore, they satisfy
$$\prod_{i=1}^N{\tanh\left({z^{+}_k-\t_i\over{2}}+i|\a|\pi\right)\over{
\tanh\left({z^{+}_k-\t_i\over{2}}\right)}}=-F,\qquad
\prod_{i=1}^N{\tanh\left({z^{-}_k-\t_i\over{2}}-i|\a|\pi\right)\over{
\tanh\left({z^{-}_k-\t_i\over{2}}\right)}}=-F.\eqn\ffmxiv$$
The solutions of these equations can be written in terms of real variables
$x_k$ in the following way:
$$\eqalign{&z^{+}_k=x_k-i|\a|\pi,\qquad
x_k-i|\a|\pi+i\pi,\cr
&z^{-}_k=x_k+i|\a|\pi,\qquad
x_k+i|\a|\pi-i\pi,\cr}\eqn\ffmxv$$
where a real number $x_k$ satisfies
$$\prod_{i=1}^N{\tanh\left({x_k-\t_i\over{2}}-{i|\a|\pi\over{2}}\right)\over{
\tanh\left({x_k-\t_i\over{2}}+{i|\a|\pi\over{2}}\right)}}=-F.\eqn\ffmxvi$$

Out of four possible choices of $z^{\pm}_k$ for $k=1,\ldots,N$ from Eq.\ffmxv,
only two choices are allowed.
This can be understood easily if one considers the limit of $|\a|\to 0$.
The Boltzmann weights are either $+1$ or $-1$ from Eq.\ffmv.
This means the transfer matrix is just a constant matrix without any dependence
on the rapidities.
Now from Eqs.\ffmxiii\ and \ffmxvi, the only possibility for the eigenvalues of
the transfer matrix to be independent of $\t_i$ is when $\{x_k\}=\{\t_i\}$
and when $z_k^{+}-z_k^{-}=\pm i\pi$ for all $k$.
For example, if one choose $(z^{+}_k,z^{-}_k)=(\t_k-i|\a|\pi,\t_k+i|\a|\pi)$
pair for some $k$ as $|\a|\to 0$, the eigenvalue will get term like
$\tanh\left({u-\t_k\over{2}}\right)$. Obviously, this eigenvalue should be
excluded for the constant transfer matrix.
This leaves only two choices for the zeroes:
$$(z_k^{+},z_k^{-})=(x_k-i|\a|\pi,x_k+i|\a|\pi-i\pi)
\quad{\rm or}\quad (x_k-i|\a|\pi+i\pi,x_k+i|\a|\pi).\eqn\ffmxvii$$

{}From the product form of Eq.\ffmxii, one notices that if we  choose one pair
of zeroes in Eq.\ffmxvii\ the other pair become zeroes of $\L(u+i\pi)$.
Since one can choose the zeroes between the two possibilities for each $k$
($k=1,\ldots,N$), we can construct $2^N$ different eigenvalues in this way.
Also, one can convince that Eq.\ffmxiii\ satisfies Eq.\ffmxii\ because if we
dividing the RHS of Eq.\ffmxii\ with $\L(u)\L(u+i\pi)$ using Eq.\ffmxiii\
the final expression has no poles and zeroes while it is bounded.
This means the ratio should be a constant.

\topinsert
\input pictex
\beginpicture
\setcoordinatesystem units <1pt,1pt>
\linethickness 4.0pt
\setlinear
\plot 30 0 330 0 /
\plot 180 100 180 -60 /
\plot 300 105 300 90 /
\plot 300 90 315 90 /
\plot 180 100 177 93 /
\plot 180 100 183 93 /
\plot 330 0 323 3 /
\plot 330 0 323 -3  /
\put {$\theta$} at 310 100
\put {$\circ$} at 30 70
\put {$\bullet$} at 60 70
\put {$\circ$} at 90 70
\put {$\bullet$} at 120 70
\put {$\circ$} at 150 70
\put {$\bullet$} at 180 70
\put {$\bullet$} at 210 70
\put {$\circ$} at 240 70
\put {$\bullet$} at 270 70
\put {$\circ$} at 300 70
\put {$\bullet$} at 330 70
\put {$\bullet$} at 30 -30
\put {$\circ$} at 60 -30
\put {$\bullet$} at 90 -30
\put {$\circ$} at 120 -30
\put {$\bullet$} at 150 -30
\put {$\circ$} at 180 -30
\put {$\circ$} at 210 -30
\put {$\bullet$} at 240 -30
\put {$\circ$} at 270 -30
\put {$\bullet$} at 300 -30
\put {$\circ$} at 330 -30
\put {$i\pi-i|\alpha|\pi$} at 190 80
\put {$-i|\alpha|\pi$} at 160 -40
\endpicture
\vskip 1cm
\centerline{\noindent{\bf Figure 1.}\quad{
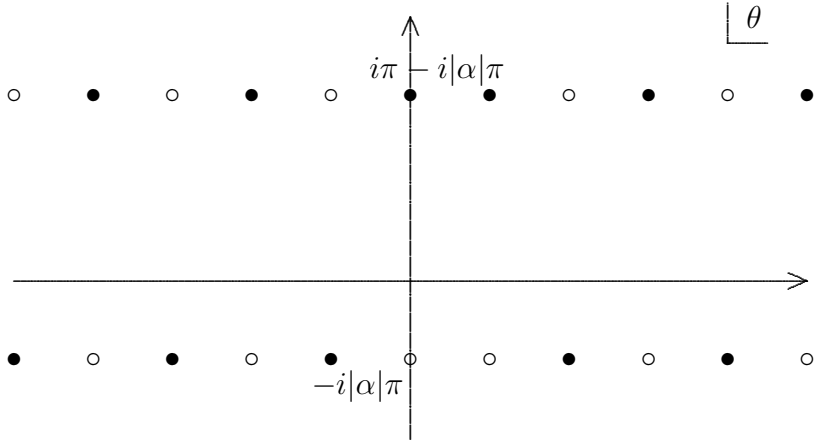
\tenpoint The zeroes $z_k^{+}$ on the complex $\t$ plane.}}
\vskip 1cm
\endinsert

Using all these results, the eigenvalues are compactly expressed by
$$\L(u)_{\ep_1,\ldots,\ep_N}={\rm Const.}\left[{\prod_{k=1}^{N}\l_{\ep_k}
(u-x_k)\over{\prod_{i=1}^{N}\sinh(u-\t_i)}}\right],
\qquad\ep_k=\pm 1,\eqn\ffmxviii$$
with
$$\l_{\ep}(\t)=\sinh\left({\t\over{2}}+\ep{i|\a|\pi\over{2}}\right)
\cosh\left({\t\over{2}}-\ep{i|\a|\pi\over{2}}\right).\eqn\ffmxix$$
$\ep=+1$ corresponds to the first choice in Eq.\ffmxvii\ and $\ep=-1$
the second. (See Fig.1)
The real zeroes $x_k$ are determined by Eq.\ffmxvi.

\section{Nondiagonal TBA}

{}From Eq.\eqPBCi\ and Eq.\ffmxviii, the PBC equation becomes
$$\half e^{im\sinh\t}\prod_{i=1}^{N}\left[{Y(\t-\t_i)\over
{\sinh(\t-\t_i)}}\right]\prod_{k=1}^{N}\l_{\ep_k}(\t|x_1,\ldots,x_N)=1,
\eqn\tbaxv$$
and the constraint equation \ffmxvi\ in the limit
$N,L\to\infty$ are expressed by introducing the densities
$\r$ for the allowed states, $\r^{1}$ for the occupied states,
$P_{+}$ for $\ep=+1$ zero-state, and $P_{-}$ for $\ep=-1$.
In terms of these densities, one gets
$$\eqalign{
&2\pi\r(\t)=m\cosh\t\cr
&+\int d\t' [\r^1(\t')\P_{Y}(\t-\t')+ P_{+}(\t')\P_{+}(\t-\t')
            +P_{-}(\t')\P_{-}(\t-\t')],\cr
&2\pi P(\t)=\int d\t' \r^1(\t')\P_{T}(\t-\t'),\cr}\eqn\tbaxvi$$
where
$$\eqalign{
&\P_{Y}(\t)={\pd\over{\pd\t}}{\rm Im}\ln
\left[{Y(\t)\over{\sinh\t}}\right],\qquad
\P_{\pm}(\t)={\pd\over{\pd\t}}{\rm Im}\ln\l_{\pm}(\t),\cr
&\P_{T}(\t)={1\over{i}}{\pd\over{\pd\t}}\ln\left[{\tanh\left({\t\over{2}}-
{i|\a|\pi\over{2}}\right)\over{\tanh\left({\t\over{2}}+
{i|\a|\pi\over{2}}\right)}}\right].\cr}\eqn\tbaxvii$$

Using Eq.\ffmxix\ and $\l_{-}=(\l_{+})^*$, one can easily show that the kernels
are related by
$$\P_{T}(\t)=2\P_{+}(\t)=-2\P_{-}(\t)={1\over{i}}{\pd\over{\pd\t}}\ln
\left[{\sinh\t-i\sin|\a|\pi\over{\sinh\t+i\sin|\a|\pi}}\right],\eqn\tbaexi$$
which is nothing but the kernel of the sinh-Gordon model.
We will denote this kernel by $\P$.
Also, we can eliminate $P_{-}$ from the first equation of \tbaxvi\ using the
second equation and $P=P_{+}+P_{-}$ to rewrite it as
$$2\pi\r(\t)=m\cosh\t+\int d\t'\left[\r^1(\t')
[\P_{Y}-\half\P*\P](\t-\t')+
P_{+}(\t')\P(\t-\t')\right],\eqn\tbaxviii$$
where we introduce a convolution defined by
$$[f*g](\t)=\int_{-\infty}^{\infty}{d\t'\over{2\pi}}f(\t-\t') g(\t').$$

Eqs.\tbaxvi\ and \tbaxviii\ have the same form as those met in diagonal TBA
like Eq.\tbav,
except that the second equation in Eq.\tbaxvi\ has no driving term.
The TBA equations, therefore, can be written down as before,
$$\eqalign{mR\cosh\t&=\ep(\t)+\left([\P_{Y}-\half\P*\P]*\ln[1+e^{-\ep}]\right)
(\t)+\left(\P*\ln[1+e^{-{\cal E}}]\right)(\t)\cr
0&={\cal E}(\t)+\left(\P*\ln[1+e^{-\ep}]\right)(\t),\cr}\eqn\tbaxix$$
in terms of the pseudoenergies $\ep$ and ${\cal E}$ defined by
$${\r^{1}(\t)\over{\r(\t)}}={e^{-\ep(\t)}\over{1+e^{-\ep(\t)}}},\qquad
{P_{+}(\t)\over{P(\t)}}={e^{-{\cal E}(\t)}\over{1+e^{-{\cal E}(\t)}}}.
\eqn\tbaxx$$

\section{Central Charges of the SShG model}

In the UV limit, Eq.\tbaxix\ reduces to simple algebraic equations of
the variables $x=\exp[-\ep(0)]$, $X=\exp[-{\cal E}(0)]$ as argued before.
For the SShG model, the algebraic equations become
$$x=(1+x)^a(1+ X)^b,\qquad X=(1+x)^b,\eqn\tbaxxi$$
with
$$a=\int_{-\infty}^{\infty}{d\t\over{2\pi}}\left(\P_{Y}-\half\P*\P\right)(\t),
\qquad b=\int_{-\infty}^{\infty}{d\t\over{2\pi}}\P(\t).\eqn\tbaxxii$$
It is not  difficult to compute these exponents using Eqs.\ffmxix\ and
\tbaxvii,
$$\int_{-\infty}^{\infty}{d\t\over{2\pi}}\P_{Y}(\t)=\half,\qquad
\int_{-\infty}^{\infty}{d\t\over{2\pi}}\P(\t)=1,$$
and
$$\int_{-\infty}^{\infty}{d\t\over{2\pi}}[\P*\P](\t)=
\left[\int_{-\infty}^{\infty}{d\t\over{2\pi}}\P(\t)\right]^2=1,$$
$$a=0\qquad{\rm and}\qquad b=1.\eqn\tbaxxiii$$

Using these values, the solution of Eq.\tbaxxi\ can be found easily as
$$x=\infty\qquad {\rm and}\qquad X=\infty.\eqn\tbaxxiii$$
One also needs the psuedo-energies as $\t\to\infty$.
Since the mass term for $\ep(\t)$ is non-zero, $\ep$ diverges as $\t\to\infty$,
and $y=\exp[-\ep(\infty)]=0$.
Then, from the second equation of \tbaxix, $Y=\exp[-{\cal E}(\infty)]=1$.

Using all these values in Eq.\tbaxiv\ the UV central charge becomes
$${6\over{\pi^2}}\left[{\cal L}(1)+{\cal L}(1)-{\cal L}
\left(\half\right)\right]={3\over{2}},$$
with $\D_{\rm min}=\ol{\D}_{\rm min}=0$ for the SShG model.
This is correct UV central charge of the SShG model with a boson and a fermion.

\section{Central Charges of the SUSY Yang-Lee Model}

As explained in the previous section, one can truncate all solitons from
the SSG multi-soliton Hilbert space to have only breathers.
In particular, for the coupling constant $\a={1\over{3}}$,
only the lightest breather and its superpartner can exist in the spectrum
with the $S$-matrix given in Eq.\Sahn\ [\Ahn].
This is the SYL model perturbed by the least relevent operator.

The fundamental difference from the SShG $S$-matrix is that
because of $\a>0$ $S_0(\t)$ is no more CDD factor. It has a pole which
is identified with the particle itself.
If we denote the particles as $B$ and $F$, the bootstrap relations are
$$BB\ (FF)\to B\to BB\ (FF),\qquad BF\ (FB)\to F\to BF\ (FB).$$
Except this difference, all the TBA analysis of the SShG model is equally
applicable to the SYL model.

The SYL conformal theory can be constructed as a coset CFT given by
$${SU(2)_K\otimes SU(2)_L\over{SU(2)_{K+L}}},\quad K=2\quad{\rm and}\quad
L+2={2\over{3}}.$$
The central charge of the model is $C=-{21\over{4}}$.
Due to the nonunitarity of the model, the lowest conformal dimension is not
zero.
To determine the lowest conformal dimension of the model, we
refer to the general formula of the general coset theories.
The primary fields of the coset theory have the conformal dimensions
given by the following formula [\CLT]:
$$\D^{l}_{r,s}={l(l+2)\over{4(K+2)}}-{l^2\over{4K}}+{(rp'-sp)^2-(p'-p)^2
\over{4Kpp'}},\quad L+2={p\over{q}},\quad p'=p+Kq,\eqn\SYLi$$
with the restrictions,
$$0\le l\le K,\quad1\le r\le p-1,\quad 1\le s\le p'-1,\quad
l=|r-s\ {\rm mod}\ 2K|.$$

For the SYL model with the values of $K=2$, $p=2$, $p'=8$,
one finds that only $r=1$,
$l=0,1,2$, and $s=1,2,3,5,6,7$ are allowed.
The minimal conformal dimension  arises when $(8-2s)^2$ is minized, i.e. with
$s=3$.  Therefore, the minimal conformal dimension of the SYL model
is $\D_{\rm min}=\D^{2}_{1,3}=-{1\over{4}}$.

Now we compute the Casimir energy of the model using TBA.
Notice that the only change from the SShG TBA is that the kernel $\P_{Y}$
in Eq.\tbaxix\ gets extra factor  $-{\pd\over{\pd\t}}\ln S_0(\t)$
due to $S_0$ in Eq.\Sahn.
With $\a>0$ this introduces extra $-1$ in the exponent $a$ in
Eq.\tbaxxi\ to make $a=-1$.
With this change the algebraic equations now become
$$x={1+X\over{1+x}},\qquad X=1+x,\eqn\SLYLii$$
and the solutions are $x=\sqrt{2}$ and $X=1+\sqrt{2}$.
Inserting this into Eq.\tbaxiv, one can get
$${6\over{\pi^2}}\left[{\cal L}\left({\sqrt{2}\over{1+\sqrt{2}}}\right)+
{\cal L}\left({1+\sqrt{2}\over{2+\sqrt{2}}}\right)-{\cal L}\left(\half\right)
\right]={3\over{4}},$$
which is exactly  $C-24\D_{\rm min}$ as required.

\endpage


\chapter{Form Factors of the Supersymmetric Theories}

In this section, we derive FFs of the SShG model from the $S$-matrix
of the model.
Starting with the axioms for the FFs to satisfy, we write two-point FFs
in terms of two factors. The first factor is determined by the properties
of operators and contents of poles while the second factor is completely
determined by the eigenvalues of the $S$-matrix.
For the SShG model, we compute one-point FFs first to fix
overall normalization.
Then, two-point FFs will be derived using the Watson equations and SUSY
relations of the FFs.

\section{Form Factor Axioms}

We start with axioms of the FFs [\Smibook].
Denoting $\ket{a(\t)}$ as an on-shell particle state of type $a$ with a
rapidity $\t$, a matrix element of an Hermitian
operator ${\cal O}$ between vacuum and
in-coming states can be expressed by
$$F^{\cal O}_{a_1,\ldots,a_n}(\t_1,\ldots,\t_n)=\bra{0}{\cal O}(0)
\ket{a_1(\t_1),\ldots,a_n(\t_n)}_{\rm in},\eqn\ffi$$
with a normal ordering of rapidities, $\t_1>\t_2>\ldots>\t_n$.
This FF should satisfy the following axioms:

1. $F$ is analytic in each variable $\t_{ij}=\t_i-\t_j$ inside the strip of
$0\le {\rm Im}[\t_{ij}]\le 2\pi$ except for simple poles.
All other type of FFs can be reduced to the standard form like
$$\eqalign{& ^{\rm out}\bra{b_1(\t'_1),\ldots,b_m(\t'_m)}{\cal O}(0)
\ket{a_1(\t_1),\ldots,a_n(\t_n)}_{\rm in}\cr
=&C^{b_1 b'_1}\cdots  C^{b_m b'_m}F_{b'_1,\ldots,b'_m,a_1,\ldots,a_n}
(\t'_1+i\pi,\ldots,\t'_m+i\pi,\t_1,\ldots,\t_n),\cr}\eqn\ffii$$
with the charge conjugation matrix $C$ arising in the crossing process.

2. Due to the normal ordering, any exchange of two rapidities should involve a
scattering process;
$$\eqalign{&F_{a_1,\ldots,a_i,a_{i+1},\dots,a_n}(\t_1,\ldots,\t_i,\t_{i+1},
\ldots,\t_n)\cr
=&(-1)^{f_{a_i} f_{a_{i+1}}} S_{a_i a_{i+1}}^{a'_i a'_{i+1}}
(\t_i-\t_{i+1}) F_{a_1,\ldots,a_{i+1},a_{i},\dots,a_n}
(\t_1,\ldots,\t_{i+1},\t_{i},\ldots,\t_n),\cr}\eqn\ffiii$$
with the phase factor arising from the exchange of particles
($f_a=0$ for a boson or $1$ for a fermion).
\foot{We do not need  this phase factor if we use Eq.\Sahn\ as
the $S$-matrix instead of Eq.\Sshawit\ since all particles are bosonic.}

3. Relativistic invariance dictates
$$F_{a_1,\ldots,a_n}(\t_1+\L,\ldots,\t_n+\L)=e^{s\L}
F_{a_1,\ldots,a_n}
(\t_1,\ldots,\t_n),\eqn\ffiv$$
where $s$ is the spin of the operator ${\cal O}$.

4. A $2\pi i$ translation of one of rapidity is equivalent to
$$F_{a_1,\ldots,a_n}(\t_1,\ldots,\t_n+2\pi i)=(-1)^{f_{a_n}\sum f_{a_i}}
F_{a_n,a_1,\ldots,a_{n-1}}(\t_n,\t_1,\ldots,\t_{n-1}).\eqn\ffv$$
There are two origins of the poles. One is the annihilation pole and the other
corresponds to the bound states.
Existence of the poles give extra constraints on the FFs.

5. The annihilation pole arises in the channel of two incoming particles
which are related to each other by the charge conjugation at $\t=i\pi$.
This gives
$$\eqalign{&i{{\rm Res}\atop{\t'\to \t}}
F_{a_1,\ldots,a_n,\ol{a},a}(\t_1,\ldots,\t_n,\t'+i\pi,\t)\cr
=&\left[{\bf 1}-{\cal T}_{aa}(\t|\t_1,\ldots,\t_n)\right]_{a_1,\ldots,a_n}
^{a'_1,\ldots,a'_n}F_{a'_1,\ldots,a'_n}(\t_1,\ldots,\t_n),\cr}\eqn\ffvi$$
using the monodromy matrix defined in Eq.\eqPBCi.
This equation relates a $n+2$-point to a $n$-point function.

6.  The bound state  pole of a $S$-matrix with a residue
$$i{{\rm Res}\atop{\t\to \t_c}}S_{ab}^{a'b'}
(\t)=-\G_{ab}^{c}\G_{c}^{a'b'},\eqn\ffvii$$
the FF should satisfy
$$i{{\rm Res}\atop{\t_a-\t_b\to \t_c}}F_{a_1,\ldots,a_n,a_a,a_b}
(\t_1,\ldots,\t_n,\t_a,\t_b)=\G_{ab}^{c}F_{a_1,\ldots,a_n,a_c}
(\t_1,\ldots,\t_n,\t_c).\eqn\ffviii$$

\section{Two-point Form Factors}

The general axioms can be much simplified for two-point FFs.
It can be written in terms of two factors like,
$$F_{a_1 a_2}(\t_1,\t_2)=K_{a_1 a_2}(\t_1,\t_2) F^{\rm min}_{a_1 a_2}
(\t_1-\t_2),\eqn\ffix$$
where $F^{\rm min}$ satisfies the Watson equation without any pole,
$$F^{\rm min}_{a_1 a_2}(\t)= F^{\rm min}_{a_1 a_2}(-\t)
S^{a'_1 a'_2}_{a_1 a_2}(\t),\quad F^{\rm min}_{a_1 a_2}(i\pi-\t)=
F^{\rm min}_{a_1 a_2}(i\pi+\t),\eqn\ffx$$
and the prefactor $K_{a_1 a_2}(\t_1,\t_2)$ has all the required poles and
operator dependence.
Note we omitted a phase factor from the fermion exchange operator treating
all particles as bosonic.

$F^{\rm min}$ can be determined from the following steps:
In the basis which diagonalizes the $S$-matrix, Eq.\ffx\ becomes a simple
functional relation.
Then, using an integral representation for the $i$-th eigenvalue of the
$S$-matrix, one finds
$$S_{i}(\t)=\exp\left[\int_{0}^{\infty}{dt\over{t}}f_{i}(t)
\sinh{\t t\over{\pi i}}\right]\longrightarrow
F_{i}^{\rm min}=\exp\left[\int_{0}^{\infty}{dt\over{t}}{f_{i}(t)
\over{\sinh\t}}\sin^2 {\wh{\t}t\over{2\pi}}\right],\eqn\ffxi$$
where $\wh{\t}=i\pi-\t$.
Rotating back to the original on-shell two-particle states, one finds the
$F^{\rm min}$.

The function $K(\t_1,\t_2)$ should be determined  by the other  axioms.
Eq.\ffv\ requires it be a symmetric function of the rapidities if $a_1=a_2$ and
has a $i\pi$-pole if $\ol{a}_1=a_2$.
The asymptotic behaviour under the rapidity translation is related to the spin
of the operator by Eq.\ffiv.
The overall normalization of the FFs is fixed by the one-point function.

\section{One-Point Form Factor of the SShG model}

We work out one-point function of the SShG model to fix normalization of the
general FFs.  From the Fourier transformation of the elementary fields
$$\eqalign{
\p({\bf x})&=\int{dk\over{2\pi}}{1\over{\sqrt{2}k_0}}
\left[b_ke^{i{\bf k}\cd{\bf x}}+b^{\dagger}_ke^{-i{\bf k}\cd{\bf x}}\right]\cr
\psi({\bf x})&=\int{dk\over{2\pi}}{1\over{k_0}}
\left[f_k u(k)e^{i{\bf k}\cd{\bf x}}+f^{\dagger}_k v(k)e^{-i{\bf k}\cd
{\bf x}}\right]\cr}\eqn\ffxii$$
with $u^*=v$ and with the commuation relations
$$\left\{f_k,f^{\dagger}_{k'}\right\}=2\pi  k_0 \d(k-k'),\qquad
\left[b_k,b^{\dagger}_{k'}\right]=2\pi  k_0 \d(k-k').\eqn\ffxiii$$

On-shell SUSY is determined from the SUSY transformation of the elementary
fields, Eq.\susyi\ and $\ket{b(\t)}=b^{\dagger}(\t)\ket{0}$ and
$\ket{f(\t)}=f^{\dagger}(\t)\ket{0}$:
$$\eqalign{
Q_1\ket{f(\t)}&=-i{P_{+}\over{\sqrt{2}v_1(\t)}}\ket{b(\t)}\qquad
Q_1\ket{b(\t)}=i\sqrt{2}  v_1(\t)\ket{f(\t)}\cr
Q_2\ket{f(\t)}&=i{P_{-}\over{\sqrt{2}v_2(\t)}}\ket{b(\t)}\qquad
Q_2\ket{b(\t)}=i\sqrt{2}  v_2(\t)\ket{f(\t)}.\cr}\eqn\ffxiv$$

{}From $0=\bra{0}Q_{\a}\left[\p(0)\ket{f(\t)}\right]$, one gets
$$Q_{\a}\ket{f(\t)}=-i\sqrt{2}u_{\a}(\t)\ket{b(\t)},\eqn\ffxv$$
and comparing  this with Eq.\ffxiv, one can find the spinors
$$\eqalign{
&v_1(\t)=-i\sqrt{{m\over{2}}} e^{\t/2},\qquad
v_2(\t)=\sqrt{{m\over{2}}} e^{-\t/2}.\cr}$$
This gives the SUSY transformation of on-shell states given above in
Eq.\onshellsusy.
Combining this and Eq.\ffxii, we fix the one-point function as follows:
$$\bra{0}\p(0)\ket{b(\t)}={1\over{\sqrt{2}}},\qquad
\bra{0}\psi_{\a}(0)\ket{f(\t)}=\sqrt{{m\over{2}}}
\pmatrix{ ie^{\t/2}\cr e^{-\t/2}\cr}.\eqn\ffxvi$$

\section{Two-point form factors of the SShG model}

We  compute two-point  FFs of the trace of energy-momentum tensors
and their SUSY counterparts, $\T$ and $\T_{\rm F}$ ($\ol{\T}_{\rm F}$),
given in Eqs.\eqTT,\eqsTT.
These operators are of particular interest for their role in the $C$-theorem.
First, we derive SUSY relations between the FFs using Eq.\eqTTsusy.\hfil\break
{}From $0=\bra{0}Q_{\a}[{\cal O}\ket{a_1(\t_1)a_2(\t_2)}]$,
one finds a relation
$$\bra{0}Q_{\a}[{\cal O}]\ket{a_1(\t_1)a_2(\t_2)}=-(-1)^{F({\cal O})}
\bra{0}{\cal O}\ket{Q_{\a}[a_1(\t_1)a_2(\t_2)]},\eqn\ffxvii$$
with $F$ is $1$ for fermionic and $0$ for bosonic.
This gives the following relations between the FFs:
$$\eqalign{&F^{\T}_{bb}=i{\sqrt{m}\over{2}}\left[\sqrt{x_1}
F^{\ol{\T}_{\rm F}}_{fb}+\sqrt{x_2}F^{\ol{\T}_{\rm F}}_{bf}\right],\quad
F^{\T}_{ff}=i{\sqrt{m}\over{2}}\left[\sqrt{x_1}F^{\ol{\T}_{\rm F}}_{bf}
-\sqrt{x_2}F^{\ol{\T}_{\rm F}}_{fb}\right],\cr
&F^{\T}_{bb}={\sqrt{m}\over{2}}\left[{1\over{\sqrt{x_1}}}
F^{\T_{\rm F}}_{fb}+{1\over{\sqrt{x_2}}}F^{\T_{\rm F}}_{bf}\right],\quad
F^{\T}_{ff}={\sqrt{m}\over{2}}\left[{-1\over{\sqrt{x_1}}}F^{\T_{\rm F}}_{bf}
+{1\over{\sqrt{x_2}}}F^{\T_{\rm F}}_{fb}\right],\cr}\eqn\ffxviii$$
where each FF is a function of $\t_1$ and $\t_2$ or of
$x_i=e^{\t_i}$.

\subsection{A special case of $\a\to 0$}

It is useful to consider a case of $\a\to 0$ where the $S$-matrix is of
diagonalized form of $(1,-1,1,1)$ from Eq.\Sahn.
Let's compute $F^{\T}_{bb}$ and $F^{\T}_{ff}$.
In terms of the  solution of the Watson equation,
\foot{
This solution is not unique in the sense that one can multiply any even
function of $\t$ satisfying $f(i\pi+\t)=f(i\pi-\t)$.
If we include these functions in the prefactor $K$, one can define
$F^{\rm min}$ uniquely.}
$$F^{\rm min}_{bb}=1,\qquad F^{\rm min}_{ff}=\sinh{\t\over{2}},
       \eqn\ffxix$$
the FFs can be written as
$$F^{\T}_{bb}(x_1,x_2)=K_{bb}(x_1,x_2),\qquad{\rm and}\qquad
F^{\T}_{ff}(x_1,x_2)=K_{ff}(x_1,x_2)\sinh{\t\over{2}}.\eqn\ffxx$$
Since $K$ should have the $i\pi$ pole (or at $x_1=-x_2$) and
$K(\t_1+\L,\t_2+\L)=K(\t_1,\t_2)$ because the spin of $\T$ is zero,
we can find
$$F^{\T}_{bb}(x_1,x_2)=2\pi m^2,\qquad{\rm and}\qquad
F^{\T}_{ff}(x_1,x_2)=2\pi m^2\sinh{\t\over{2}}.\eqn\ffxxi$$
Here we fixed the normalization factor as $\pi m^2$ by comparing with the
perturbative computation using Eqs.

After finding these, one can derive the other FFs simply using
Eq.\ffxviii\ as follows:
$$F^{\T_{\rm F}}_{bf}(x_1,x_2)=2\pi m^{3/2}\sqrt{x_2},\qquad
F^{\ol{\T}_{\rm F}}_{bf}(x_1,x_2)=2\pi m^{3/2}{-i\over{\sqrt{x_2}}}.
\eqn\ffxxii$$
One can check that the spins of $\T_{\rm F}$ and $\ol{\T}_{\rm F}$ can be
found correctly as $\pm\half$ under the rapidity translation.
Also one can check that
these FFs are consistent with pertrubative computation.

\subsection{For General $\a$}

For general cases, we should diagonalize the $S$-matrix first.
Whether we use Eq.\Sshawit\ with the phase factor in Eq.\ffiii\ or
Eq.\Sahn\ without such factor, we find the eigenvalues of $F=0$ sector
($bb$ and $ff$) are complicated and hard to find the integral representations.
Instead, we consider $F=-1$ ($bf$ and $fb$) sector first.
The $S$-matrix is easily diagonalized by the eigenvectors
$$\ket{+}={1\over{\sqrt{2}}}\left(\ket{b_1 f_2}+\ket{f_1 b_2}\right),
\qquad\ket{-}={1\over{\sqrt{2}}}\left(\ket{b_1 f_2}-\ket{f_1 b_2}\right),
\eqn\ffxxiii$$
with eigenvalues
$$S_{+}(\t)=\exp\left[\int_{0}^{\infty}{dt\over{t}}f_{+}(t)
\sinh{\t t\over{\pi i}}\right],\qquad S_{-}(\t)=-\exp
\left[\int_{0}^{\infty}{dt\over{t}}f_{-}(t)\sinh{\t t\over{\pi i}}\right],
\eqn\ffxxiv$$
with
$$f_{\pm}(t)={(1-\cosh t)\left(1+\cosh((1-2|\a|)t)\right)\over{\sinh^2 t}}
\pm{\cosh((1-2|\a|)t)\over{\cosh t}}.\eqn\ffxxv$$

{}From these integrals, $F^{\rm min}$ in the basis
of $\ket{+}$ and $\ket{-}$ can be obtained as
$$F^{\rm min}_{+}=
\exp\left[\int_{0}^{\infty}{dt\over{t}}{f_{+}(t)\over{\sinh t}}
\sin^2{\wh{\t}t\over{2\pi}}\right],\quad
F^{\rm min}_{-}=\cosh{\wh{\t}\over{2}}\exp\left[\int_{0}^{\infty}{dt\over{t}}
{f_{-}(t)\over{\sinh t}}\sin^2{\wh{\t}t\over{2\pi}}\right],\eqn\ffxxvi$$
where we chose a normalization such that
$$F_{\pm}\to 1\qquad{\rm as}\qquad \a\to 0.\eqn\ffxxvii$$
For numerical computations, we list expressions of $F^{\rm min}_{\pm}$
which converge fast in the Appendix B.

Now we consider FFs of $\T_{\rm F}$ in the following form:
$$F^{\T_{\rm F}}_{+}(x_1,x_2)=K_{+}(x_1,x_2)F^{\rm min}_{+},\quad
F^{\T_{\rm F}}_{-}(x_1,x_2)=K_{-}(x_1,x_2)F^{\rm min}_{-},\eqn\ffxxviii$$
and similarly for $\ol{\T}_{\rm F}$ in terms of $\ol{K}_{\pm}$.
These $K_{\pm}$ and $\ol{K}_{\pm}$ can be determined from the spins
of the operators and
symmetric properties of the states under the exchange, $\ket{+}\to\ket{+}$ and
$\ket{-}\to-\ket{-}$ under $x_1\leftrightarrow x_2$, as follows:
$$K_{\pm}=A(\sqrt{x_1}\pm\sqrt{x_2}),\quad
\ol{K}_{\pm}=B\left({1\over{\sqrt{x_1}}}\pm {1\over{\sqrt{x_2}}}\right),
\eqn\ffxxix$$
where  constants $A,B$ can be determined by taking $\a\to 0$ limit
and comparing with Eq.\ffxxii.

Now rotating back to the on-shell states one can find
$$\eqalign{
&F^{\T_{\rm F}}_{bf}(x_1,x_2)=2\pi m^{3/2}\left[\sqrt{x_1}
{(F^{\rm min}_{+}-F^{\rm min}_{-})\over{2}}+
\sqrt{x_2}{(F^{\rm min}_{+}+F^{\rm min}_{-})\over{2}}\right],\cr
&F^{\ol{\T}_{\rm F}}_{bf}(x_1,x_2)=-2\pi i m^{3/2}\left[{1\over{\sqrt{x_1}}}
{(F^{\rm min}_{+}-F^{\rm min}_{-})\over{2}}+
{1\over{\sqrt{x_2}}}{(F^{\rm min}_{+}+F^{\rm min}_{-})\over{2}}\right].\cr}
\eqn\ffxxx$$
Also from Eq.\ffxviii\ one can obtain other FFs
$$\eqalign{
&F^{\T}_{bb}(x_1,x_2)=2\pi m^{2}\left[{(F^{\rm min}_{+}+F^{\rm
min}_{-})\over{2}}
+{(F^{\rm min}_{+}-F^{\rm min}_{-})\over{2}}\cosh{\t\over{2}}\right],\cr
&F^{\T}_{ff}(x_1,x_2)=2\pi m^{2}
{(F^{\rm min}_{+}+F^{\rm min}_{-})\over{2}}\sinh{\t\over{2}}.\cr}
\eqn\ffxxxi$$
We checked these FFs using the first order perturbative computations.
FFs for other component of energy-momentum  tensor can be written down
by just multiplying $P_{+}/P_{-}$ to the above FFs.

\endpage


\chapter{Spectral Representation of $C$-Theorem}

We compute the UV central charge of the SShG model using the two-point FFs
computed in the previous section
and the spectral representation of the $C$-Theorem.
This provides a consistency check for the FFs and shows the fast
convergence of the FFs expansions of correlation functions.

\section{Spectral Sum Rule}

The $C$-theorem, first introduced by A.B. Zamolodchikov,
plays an important role in the study of off-critical models [\ZamC].
The $C$-function, describing a degree of freedom of the 2D models,
connects smoothly two renormalization group (RG) fixed points
as the length scale of the theory increases from UV limit to IR.
For some specific models like the perturbed minimal CFTs by the least
relevent operator with positive coefficient, the renormalization
group (RG) flow connects two RG fixed points corresponding to two
adjacent minimal CFTs [\LudCar].
This RG flow will end up at the massive point with $C=0$.

This theorem can be neatly expressed in the following integral of the two-point
correlation function of the trace of energy-momentum tensor following
Cardy [\Cardy]:
$$\eqalign{
&\D C={3\over{4\pi}}\int_{|x|>\varepsilon}d^2 x x^2\langle
\T(x)\T(0)\rangle=\int_{0}^{\infty}d\m C_1(\m,\L),\cr
&C_1(\m,\L)={6\over{\pi^2}}{1\over{\m^3}}{\rm Im}\left[\int d^2 x e^{-ip\cd x}
\langle\T(x)\T(0)\rangle\right]_{p^2=-\m^2}.\cr}\eqn\csumi$$
Expanding the correlation function in terms of intermediate on-shell states,
the spectral density function $C_1$ can be expressed in terms of the FFs by
$$C_1(\m,\L)={12\over{\m^3}}\sum_{\a}\big\vert
\bra{0}\T(0)\ket{\a}\big\vert^2\d^2(q-p_{\a}),
\eqn\csumii$$
where $p_{\a}$ is the energy-momentum vector of the multi-particle state
$\a$ and the vector $q$ is defined as $q=(\m,0)$.

For the massive theory, the sum rule of $\D C$ effectively gives the UV
central charge since $C_{\rm IR}$ vanishes.
Although one needs the infinite number of the FFs to compute it
rigorously, there are many evidences that the sum in Eq.\csumii\ converges
very fast for the massive theories [\YurZam--\CarMus].
With this observation, one can  compute the UV central charge using the
two-point FFs of $\T$ quite accurately.
In next stage, we will compute this numerically using the FFs of
the SShG model derived in the previous section.

\section{Sum Rule for the SShG model}

The two-point contribution to the sum rule becomes
$$\eqalign{C^{(2)} =&{12\over{\m^3}}\int{d\t_1 d\t_2\over{2(2\pi)^2}}
\sum_{a_1,a_2}|F^{\T}_{a_1 a_2}(\t_1,\t_2)|^2\cr
&\times\d(m\cosh\t_1+m\cosh\t_2-\m)\d(m\sinh\t_1+m\sinh\t_2)\cr
=&{3\over{8\pi^2 m^4}}\int_{0}^{\infty}{d\t\over{\cosh^4\t}}\left[
|F^{\T}_{bb}(\t,-\t)|^2+|F^{\T}_{ff}(\t,-\t)|^2\right],\cr}\eqn\csumiii$$
with the FFs given in Eq.\ffxxxi.

For the special case of $\a=0$ where the SShG model becomes free with
a boson and fermion, one can insert Eq.\ffxxi\ into Eq.\csumiii\
and using
$$\int_{0}^{\infty}{d\t\over{\cosh^{4}\t}}={2\over{3}},\qquad
\int_{0}^{\infty}d\t{\sinh^2\t\over{\cosh^4\t}}={1\over{3}},$$
one can easily find $C={3\over{2}}$.

For the generic value of $\a$ we
integrate numerically using the regularized
expressions for the $F_{\pm}$ in Appendix B.
Using these we list $\D C^{(2)}$ as for several values of the coupling
constant in Table 1.
This shows a good agreement with the UV central charge $C={3\over{2}}$.
The convergence of the SShG model seems slow
compared with the sinh-Gordon result [\FMS].
This suggests in the SShG model one arrives strong coupling region earlier
than the sinh-Gordon model as one can see from the fact that
the limit of the SShG coupling constant is
${\wh{\b}^2\over{8\pi}}={1\over{2}}$ while
${\wh{\b}^2\over{8\pi}}=1$ in the sinh-Gordon model.

\topinsert
\input tables
\bigskip
\begintable \qquad ${\beta^2\over{4\pi}}$\qquad|\qquad
$\alpha$\qquad|\qquad $\Delta \wh{C}^{(2)}$\qquad\crthick
${1\over{999}}$|$0.001$| $0.9993$\cr
${1\over{199}}$|$0.005$|$0.9953$\cr
${1\over{99}}$|$0.01$|$0.9902$\cr
${1\over{49}}$|$0.02$|$0.9800$\cr
${3\over{97}}$|$0.03$|$0.9697$\cr
${1\over{19}}$|$0.05$|$0.9495$\cr
${1\over{9}}$|$0.1$|$0.9093$\endtable
\vskip 1cm
\centerline{\noindent{\bf Table 1.}\quad{\tenpoint
The first two-particle form factor in the Sum Rule of
$\Delta\wh{C}={2\over{3}}\D C$.}}
\vskip 1cm
\endinsert
\endpage


\chapter{Conclusion}

In this paper we obtained two results on the $N=1$ SUSY integrable models.
The first one is the computation of the UV central charges from TBA method.
The nondiagonal TBA of the SShG and SYL models has been rigorously
derived from the essential observation that
the $N=1$ SUSY models can be identified with the eight vertex free fermion
models.
These TBA equations produced correct UV central charges.

The second result is two-point FFs of the SShG model using
the FF axioms.
Here the difficulty arising from the nondiagonal scattering theories has
been avoided from the SUSY relations of the FFs.
The spectral representation of the $C$-theorem showed that two-point
FFs can give good approximations in the infinite sum of the intermediate
states even in nondiagonal theories.

Our results suggest some interesting directions to proceed further.
Actually, we notice that wider class of $N=1$ scattering theories are
belonging to the eight vertex FFMs which will be reported in separate
publication [\Ahnnext].
The relationship between these SUSY models and the eight vertex FFM may have
some deep structure because
the FFMs seem to have interesting hidden symmetries [\BazStr].
In particular, it has been noticed recently that the FFMs have a hidden
quantum group symmetry [\CGLS].
It would be interesting to see how this quantum group symmetry will be
related to the $N=1$ supersymmetry in the trigonometric limit.

In this paper, we could not say much on the general FFs of the theories.
The solution of the FF bootstrap equations are very difficult and
are limited to only a few simplest diagonal theories.
We can reduce, however, the nondiagonal bootstrap equations
to the level of diagonal theories
by diagonalizing the inhomogeneous transfer matrix.
It will need some more work to solve these reduced bootstrap equations
completely.

\ack

\noindent The author wish to thank T. Eguchi, A. LeClair, G. Mussardo,
M. Peskin, F. Smirnov, P. Weisz for useful discussions.

\Appendix A{Inversion relation for the Free Fermion Model}

We follow Felderhof to diagonalize the transfer matrix of the FFM [\Felderhof].
We want to point out, first,
the difference of our derivation from the lattice model computation.
The first difference is that we want to
diagonalize the inhomogeous transfer matrix.
This difference often introduces much difficulty for the computation.
However, this difficulty can be avoided by the second difference, which is that
we are working with the FFM at the critical point.
With this advantage, we can derive the inverse matrix of the FFM transfer
matrix and, furthermore, express it using the original transfer matrix with
slight change in the rapidity $u$.

It is convenient to reexpress the Boltzman weights Eq.\ffmi\ in terms of
the $\s$-matrices,
$$R(\t)=\pmatrix{A(\t)&B(\t)\cr C(\t)&D(\t)\cr},\eqn\appci$$
$$\eqalign{&A=a_{+}\s^{+}\s^{-}+b_{+}\s^{-}\s^{+},\qquad B=d\s^{+}+c\s^{-},\cr
&C=c\s^{+}+d\s^{-},\qquad D=b_{-}\s^{+}\s^{-}+a_{-}\s^{-}\s^{+}.\cr}
\eqn\appcii$$
Then, the transfer matrix becomes
$$T(u|\t_1,\ldots,\t_N)={\rm Tr}_{2}\left[\prod_{i=1}^{N}R(u-\t_i)\right].
\eqn\appciii$$

Now we define new transfer matrix $T_1$ corresponding to new Boltzman weghts
defined by
$$a^{1}_{\pm}=-b_{\pm},\quad b^{1}_{\pm}=a_{\pm},\quad c^{1}=c,\quad
{\rm and}\quad d^{1}=-d.\eqn\appciv$$
In the same way as before, one can express $T_1$ by
$$T_1(u|\t_1,\ldots,\t_N)={\rm Tr}_{2}\left[\prod_{i=1}^{N}R_1(u-\t_i)\right],
\qquad R_1(\t)=\pmatrix{A_1(\t)&B_1(\t)\cr C_1(\t)&D_1(\t)\cr},\eqn\appcv$$
where
$$\eqalign{&A_1=-b_{+}\s^{+}\s^{-}+a_{+}\s^{-}\s^{+},\qquad
B_1=-d\s^{+}+c\s^{-},\cr
&C_1=c\s^{+}-d\s^{-},\qquad D_1=a_{-}\s^{+}\s^{-}-b_{-}\s^{-}\s^{+}.\cr}
\eqn\appcvi$$
One can check that these new Boltzman weights again satisfy the free fermion
condition Eq.\ffmii.

Next step is to show that $T T_1\propto 1$. For this purpose, we multiply two
matrices
$$T(u)T_1(u)={\rm Tr}_{2}\left[\prod_{i=1}^{N} R_i\right]{\rm Tr}_{2}
\left[\prod_{i=1}^{N} R_{1,i}\right]=
{\rm Tr}_{2\otimes 2}\left[\prod_{i=1}^{N} R_i\otimes R_{1,i}\right].
\eqn\appcvii$$
Defining the $4\times 4$ matrix $R_i\otimes R_{1,i}$ as $S_i$,
one can find a similarity transformation $S'_i=X_i S_i X^{-1}_i$ where
$S'_i$ is of triangular form.
The $X$ and $S'$ are given by
$$X=\pmatrix{0&{1\over{\sqrt{2}}}&{1\over{\sqrt{2}}}&0\cr
\cosh\p&0&0&-\sinh\p\cr -\sinh\p&0&0&\cosh\p\cr
0&{1\over{\sqrt{2}}}&-{1\over{\sqrt{2}}}&0\cr},\qquad
S'=\pmatrix{M_{+}&*&*&*\cr 0&F_{-}\s^{z}&0&*\cr 0&0&F_{+}\s^{z}&*\cr
    0&0&0&M_{-}\cr},\eqn\appcviii$$
where $M_{\pm}$, $F_{\pm}$, and $\p$ are given in Eqs.\ffmviii\ and \ffmext.
We did not specify the unnecessary non-vanishing components ($*$).

Most important observation is that $\tanh\p$ becomes just a constant for the
$N=1$ supersymmetric theory.
This means one can make all the $S_i$'s in the trace of triangular form
by the same similarity transformation $X$.
Therefore, $TT_1={\rm Tr}_{4}\prod S'_i$ and from Eq.\appcviii\
one can derive
$$\eqalign{
T(u)T_1(u)=&\Biggl[\prod_{i=1}^{N}M_{+}(u-\t_i)+\prod_{i=1}^{N}M_{-}(u-\t_i)\cr
&\left.+{F}\left(\prod_{i=1}^{N}F_{+}(u-\t_i)+\prod_{i=1}^{N}F_{-}(u-\t_i)
\right)\right],\cr}\eqn\appcix$$
with $F=\prod\s^{z}_i$ is either $1$ (bosonic) or $-1$ (fermionic).

Now, consider a  translation $u\to u+i\pi$. Under this the Boltzman weights of
the SShG model change
$$a_{\pm}\to -a_{\mp},\quad b_{\pm}\to b_{\mp},\quad c\to d,\quad{\rm and}\quad
d\to -c.\eqn\appcx$$
Again this satisfies the free fermion condition.
Now, the transfer matrix with translated rapidity can be expressed in terms of
$R_2(u-\t)=R(u+i\pi-\t)$ by,
$$T(u+i\pi)={\rm Tr}_{2}\left[\prod_{i=1}^{N}R_{2}(u-\t_i)\right],\qquad
R_{2}=\pmatrix{A_2&B_2\cr C_2&D_2\cr},\eqn\appcxi$$
where
$$\eqalign{&A_2=-a_{+}\s^{+}\s^{-}+b_{+}\s^{-}\s^{+},
\qquad B_2=-c\s^{+}+d\s^{-}=,\cr
&C_2=d\s^{+}-c\s^{-},\qquad D_2=b_{-}\s^{+}\s^{-}-a_{-}\s^{-}\s^{+}.
\cr}\eqn\appcxii$$
{}From Eq.\appcvi, one can notice that
$$A_2=-D_1,\qquad B_2=-C_1,\qquad,C_2=-B_1,\quad{\rm and}\quad D_2=-A_1.
\eqn\appcxiii$$

Considering the $R$-matrices as $2\times 2$ matrices, the $R_1$ and $R_2$ are
related by
$$R_2=-\s^{x}R_1\s^{x},$$
where $\s^{x}$ is the usual Pauli spin matrix.
This gives
$$T(u+i\pi|\t_1,\ldots,\t_N)=(-1)^{N} T_1(u|\t_1,\ldots,\t_N),\eqn\appcxiv$$
and from Eqs.\appcix\ and \appcxiv, the inversion relation Eq.\ffmvii.

\endpage

\Appendix B{Regularized Expression for the Form Factors}

For the numerical computation we can rewrite $F^{\rm min}_{\pm}$ in
Eq.\ffxx\ as follows:
$$F^{\rm min}_{\pm}(\t)
=C_{\pm}(\t)\left[\prod_{k=1}^{n}G_k(\a,\t)[H_k(\a,\t)]^{\pm 1}\right]
\exp\left[\int_{0}^{\infty}{dt\over{t}}{f^{\pm}_{n}(\a,t)
\over{\sinh t}}\sin^2{\wh{\t}t\over{2\pi}}\right],\eqn\appdi$$
with $C_{+}=1$, $C_{-}(\t)=\cosh{\wh{\t}\over{2}}$ and
$$\eqalign{
&G_k(\a,\t)={P_{k}(2|\a|+1,\t)^2 P_{k}(0,\t)^2
\over{P_{k}(1,\t)^2 P_{k}(2|\a|,\t) P_{k}(2|\a|+2,\t)}},\cr
&P_k(x,\t)=\left[
\left(1+{\wh{\t}/(2\pi)\over{(k+(1+x)/2)}}\right)
\left(1+{\wh{\t}/(2\pi)\over{(k+(1-x)/2)}}\right)\right]^{{k(k+1)\over{4}}},
\cr
&H_k(\a,\t)=\left[1+{\wh{\t}/(2\pi)\over{(2k+(2|\a|+3)/2)}}\right]^{\half}
\left[1+{\wh{\t}/(2\pi)\over{(2k-(2|\a|-1)/2)}}\right]^{\half},\cr}\eqn\appdii$$
and the exponents are given by
$$f^{\pm}_{n}(\a,t)={(1-\cosh t)\left(1+\cosh((1-2|\a|)t)\right)D_n(t)
\over{2\sinh^2 t}}\pm{\cosh((1-2|\a|)t) e^{-4nt}\over{\cosh t}},$$
with $D_n(t)=[(n+1)(n+2)-2n(n+2)+n(n+1)e^{-4t}] e^{-2nt}$.

If one choose $n=0$, this reduces to Eq.\ffxxvi. For the fast convergence, one
can increase $n$ although the final expression is independent of $n$.

\endpage

\refout

\end